# Scanning Tunneling Spectroscopy of Proximity Superconductivity in Epitaxial Multilayer Graphene

Fabian D. Natterer[1,†], Jeonghoon Ha[1,2], Hongwoo Baek[1,3], Duming Zhang[1,2], William G. Cullen[1,2], Nikolai B. Zhitenev[1], Young Kuk[3], and Joseph A. Stroscio[1,*]

[1]Center for Nanoscale Science and Technology, National Institute of Standards and Technology, Gaithersburg, MD 20899, USA
[2]Maryland NanoCenter, University of Maryland, College Park, MD 20742, USA
[3]Department of Physics and Astronomy, Seoul National University, Seoul, 151-747, Korea

We report on spatial measurements of the superconducting proximity effect in epitaxial graphene induced by a graphene-superconductor interface. Superconducting aluminum films were grown on epitaxial multilayer graphene on SiC. The aluminum films were discontinuous with networks of trenches in the film morphology reaching down to exposed graphene terraces. Scanning tunneling spectra measured on the graphene terraces show a clear decay of the superconducting energy gap with increasing separation from the graphene-aluminum edges. The spectra were well described by Bardeen-Cooper-Schrieffer (BCS) theory. The decay length for the superconducting energy gap in graphene was determined to be greater than 400 nm. Deviations in the exponentially decaying energy gap were also observed on a much smaller length scale of tens of nanometers.

When a superconductor is in electrical contact with a normal metal, the "leakage" of Cooper pairs via Andreev reflection causes the normal metal to acquire superconducting properties within a certain length scale from the interface. A resurgence of interest in proximity induced superconductivity has resulted from the ability to probe the superconductor-normal interface on the atomic scale with scanning tunneling spectroscopy (STS) [1–7]. Recent studies have probed the effects of disorder [1,5], temperature [6] and magnetic field [3], interfaces between two superconductors [4], and Josephson vortex formation [7]. Only recently has proximity induced superconductivity in graphene been reported by tunneling spectroscopy in graphene films grown on Re [8]. However, spatially resolved examination of the superconducting properties near the graphene-superconductor (GS) interface was not possible in the latter study since the entire Re surface was covered by graphene. Interest in graphene-superconductor devices stems from multiple unique properties such as ballistic Josephson junctions [9–11], gate-tunable metal-superconductor transitions [12,13], and unusual Andreev specular reflection which can occur when the electron energy is below the pairing energy of the superconductor [14]. While some of these phenomena have been observed in recent transport studies, the atomic scale STS experiments on high-quality ballistic graphene devices with well-defined GS interfaces and tunable density are challenging. Among the additional challenges are the required purity of the exposed graphene surface, possible degradation of the GS interface due to oxidation or contamination, and positioning of an STS probe within the microscopic area of a device. In this article, we introduce a novel approach that circumvents some of these challenges, namely the quality of the exposed graphene surface and of the GS interfaces, opening the path for further microscopic studies of hybrid graphene-superconductor materials.

We report the first scanning tunneling spectroscopic measurements of proximity induced superconducting graphene as a function of lateral distance from a GS boundary. Superconducting graphene was obtained by depositing aluminum films on few-layer graphene grown epitaxially on SiC(0001). The resulting aluminum films were discontinuous with full-depth voids exposing pristine graphene terraces. The tunneling spectra are well fit to BCS theory and show a coherence length in graphene of more than 400 nm. Non-monotonic variations of the superconductor order parameter were observed on a length scale much smaller than the coherence length.

The experiments were carried out in a custom built ultra-low temperature scanning tunneling microscopy (STM) system operating at 10 mK [15]. Tunneling spectroscopy was performed by measuring the tunneling differential conductance, $dI/dV$, using a lock-in detection scheme with modulation frequency of 141 Hz and a root-mean-square modulation amplitude of 15 µV added to the sample bias. The effective electron temperature was extracted from tunneling spectra on superconducting aluminum. Electrochemically polished Ir tips were used for tunneling probes. Few layer graphene films were grown on a (Si-face) SiC substrate by high temperature thermal sublimation of Si [16]. The aluminum films were grown on the graphene substrate by molecular beam epitaxy at a rate of 0.04 nm/s with the substrate temperature initially held at 20 °C for the first 4.5 nm of growth, followed by 350 °C for the rest of the film growth for the total thickness of ≈200 nm. The films were then transferred in ultra-high vacuum (UHV) and loaded into the STM module, and cooled to our base temperature of 10 mK for measurements [15]. The aluminum grew in atomically smooth films with very uniform height, but with deep voids exposing the bare graphene terraces, as shown in the atomic force microscopy (AFM) image in Fig. 1(a). The AFM measurements were done at room temperature after the sample was removed from the STM system.

The areas of exposed graphene substrate within the aluminum voids are of sub-micron length scale (see Fig. 1(a) and (b)). Inspection of the terraces between the aluminum film edges shows pristine graphene surfaces, as observed in Fig. 1(c). Multiple graphene layers are evident, firstly from the atomic step indicated in Fig. 1(b), and secondly from the moiré pattern observed in Fig. 1(c). The latter is due to a rotational misalignment between the top two layers of graphene with an angle of ≈5°. Such a rotation is commonly encountered in epitaxial graphene on SiC, and effectively decouples the top graphene layer, yielding single layer like electronic properties [17,18]. The disorder potential seen by the carrier in this decoupled graphene layer is very low as determined in our previous STM experiments [17,18].

We now focus on proximity induced superconductivity of the region measured along the dashed line indicated in Fig. 1(b). Figure 2 displays the tunneling spectra starting at the aluminum edge on the right side of the graphene terrace in Fig. 1(b), which corresponds to $x=0$. To determine the energy gaps we fit the spectra to the Maki theory, which is an extension of the BCS theory, accounting for effects of orbital depairing, the Zeeman splitting of the spin states, and spin orbit scattering [19,20]. The fitting parameters for zero field are the energy gap $\Delta$, the orbital depairing parameter $\zeta$, and the effective electron temperature $T_{\text{eff}}$. The aluminum spectrum at $x=0$ is well fit by the modified BCS theory with a superconducting gap of $(180.5 \pm 0.3)$ µeV [21]. This value is in good agreement with previous measurements of aluminum films [22]. Similarly, when we measure tunneling spectra on the exposed graphene terrace, we likewise observe well defined superconducting tunneling spectra in graphene that again obey the modified BCS theory. At a distance of 41 nm from the aluminum edge, the superconducting gap is on the order of 100 µeV, and decreases with growing distance from the aluminum edge. A normal state spectrum is, however, never observed in the present case since the area of the graphene terrace is too small for proximity effects to fully decay from all the surrounding aluminum edges.

Figure 3(a) shows the location dependent tunneling spectra in a 2D color graph along the dashed white line in Fig. 1(b). The superconducting gap (brown-yellow color) is observed throughout the spatial range along with well-defined coherence peaks (dark green) outside the gap. Each spectrum in Fig. 3(a) is fit to the modified BCS theory and the resulting superconducting gap is plotted in Fig. 3(b). An abrupt drop in the gap energy is observed next to the graphene-aluminum interface. Although such a sudden change in gap energy can be expected from theory and is observed in spatial measurements [4], we cannot rule out the effect of the finite probe tip radius sampling the several dozen nanometer high sidewall of the aluminum film edge as we approach it, giving an abrupt change in gap energy before the tip samples the graphene terrace at the bottom of the graphene-aluminum interface. Following the abrupt drop, the gap size decays and reaches a minimum around $x = 325$ nm, and then increases again as the path approaches the aluminum island on the left side of the trench. The gap vs distance data is fit to a decaying exponential (solid line in Fig. 3(b)), $\Delta \propto e^{-x/\xi}$, over the range of (325 nm $\geq x >$ 0 nm), which yields a coherence length $\xi = (429 \pm 9)$ nm. The superconducting coherence length is given by $\xi = \sqrt{\hbar D/\Delta}$, where $D$ is the diffusion constant $D = v_F l_e/2$, $l_e$ is the elastic mean free path, and $v_F$ is the Fermi velocity. Using the above value for the coherence length we can estimate the graphene elastic mean free path as $l_e = (101 \pm 4)$ nm [21].

Interestingly, deviations of 10 % to 20 % from a simple monotonic decay are observed upon a closer examination of the distance dependent gap values in Fig. 3(b). Possible explanations for these fluctuations include oscillations in the order parameter due the formation of a graphene electron resonator with the graphene terraces surrounded by aluminum islands. Oscillations in the supercurrent density has been observed in transport measurements in fabricated graphene resonators coupled to superconducting leads [9,10]. A second possibility may be due to the role of disorder in the graphene lattice. Recent spatial STM measurements of single atomic layers of Pb on silicon have shown that disorder in the 2D limit can significantly affect the superconducting order parameter on a length scale much shorter than the coherence length [5]. Indeed the graphene topography displayed in Fig. 3(c) shows small, ≈100 pm, vertical variations in the topography indicating some form of disorder. It appears as if some of the topographic changes were correlated with the energy gap oscillations (see vertical dashed lines connecting Fig. 3(b) and (c)), but the largest change in topography occurring at the graphene atomic step at $x \approx 100$ nm does not show a large variation in gap value, except for a few outlier points which resulted from poor data quality obtained right at the step edge. These topographic variations can also be seen in the STM image in the inset in Fig. 3(c), where the small variations appear as small scale bulges in the graphene lattice. These features are possibly related to lattice strain. Strain in graphene is associated with pseudomagnetic fields [23–26]. It remains an interesting theoretical problem to determine whether/how pseudomagnetic fields would affect the graphene superconducting state.

In summary, we report the first spatial mapping of proximity induced superconductivity in epitaxial graphene. Superconductivity was induced in graphene through aluminum films grown on epitaxial graphene on SiC substrates, which had exposed voids reaching down to the graphene terraces. The superconducting order parameter was determined to have a coherence length of (429 ± 9) nm [21]. The order parameter showed oscillations of ≈10 % on a much smaller length scale of tens of nanometers. These oscillations may be due to the formation of an electron resonator [9,10] inside the voids in the aluminum film or due to disorder [5] in the graphene lattice. This Al-graphene system lays the ground work for possible future scanning probe measurements on gated single layer and bilayer graphene devices to investigate superconducting graphene in greater detail as function of carrier density and Fermi wavelength.

**Acknowledgements**

F.D.N. greatly appreciates support from the Swiss National Science Foundation under project No. 158468. J.H., D.Z., and W.C. acknowledge support under the Cooperative Research Agreement between the University of Maryland and the National Institute of Standards and Technology, Center for Nanoscale Science and Technology, Grant No. 70NANB10H193, through the University of Maryland. H.B. and Y.K. are partly supported by national research foundation of Korea through Grant No. KRF-2010-00349. We would like to thank Steve Blankenship, Glen Holland, and Alan Band for technical assistance.

**Figure Captions**

Fig. 1. Growth of superconducting aluminum films on epitaxial graphene on a SiC substrate. (a) Large AFM scan of the aluminum topography showing pits between the aluminum terraces which reach down to the top graphene layer. The height scale covers a range of 198 nm from dark to bright. (b) STM image of a trench region between aluminum islands. The white arrows indicate a monolayer graphene step and the dashed line traces the position of the spectroscopy measurements in Fig. 2 and 3. (c) High resolution STM image of the graphene lattice obtained at the location marked by the red square in (b). The larger wavelength modulations are due to a moiré pattern with period of ≈2.8 nm due to rotational misalignment of ≈5° between the top and second graphene layers.

Fig. 2. Graphene superconducting tunneling spectroscopy. Differential tunneling spectra (symbols) measured at several lateral positions from an aluminum-graphene edge at $\Delta x=0$ (see Fig. 3). The solid lines are non-linear fits using the modified BCS theory by Maki [19,20] with gap energies indicated on the right of the graph [21]. The effective temperature of $T_{eff}=232$ mK, representing the residual electrical noise in the system, was determined from the best fit to the aluminum spectrum at $x=0$ and subsequently held fixed to extract the distance dependent gap width on graphene. The error in the gap energy was determined from the chi-square minimization in non-linear least square fits to the Maki theory.

Fig. 3. Proximity induced superconductivity in epitaxial graphene. (a) $dI/dV$ vs $V_b$ tunneling spectra, measured on the graphene terrace starting at an aluminum-graphene edge along the dashed lines shown in (c) and Fig. 1(b). The spectra are displayed in a color scale, where the brown color indicates a superconducting gap induced by proximity to the nearby aluminum islands. (b) The superconducting gap [21] determined by fitting the spectra in (a) to the modified BCS theory of Maki [19,20]. The dashed line is a guide to the eye to show the abrupt change in gap energy near the graphene-aluminum interface. The gap energies for (325 nm ≥ $x$ > 0 nm) are fit to an exponential decay (solid line) yielding a graphene coherence length of $\xi = (429\pm9)$ nm [21]. The error estimates in the gap energy and coherence length were determined from the chi-square minimization of the non-linear fits. (c) The graphene topographic height and STM image along the path of the spectral measurements in (a) and (b).

# References

*To whom correspondence should be addressed: joseph.stroscio@nist.gov

†Present address: IBM Research Center Almaden, 650 Harry Road, San Jose, CA 95120, USA

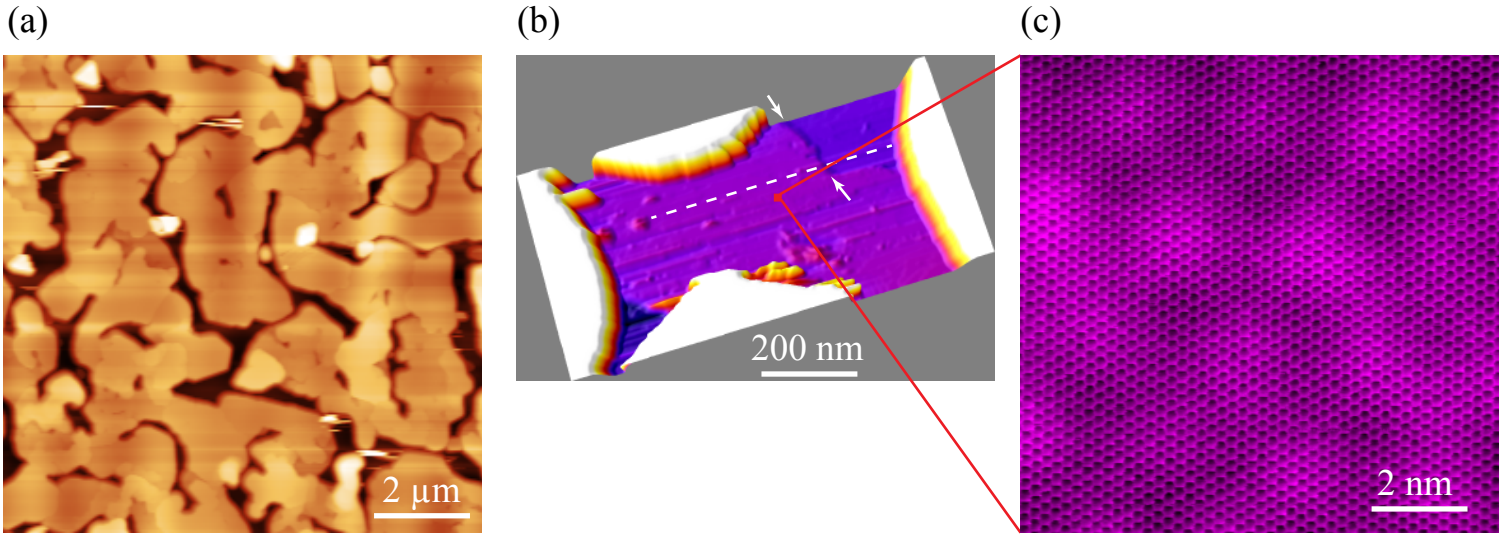

Fig. 1. Growth of superconducting aluminum films on epitaxial graphene on SiC substrates. (a) Large AFM scan of the aluminum topography showing pits between the aluminum terraces which reach down to the top graphene layer. The height scale covers a range of 198 nm from dark to bright. (b) STM image of a trench region between aluminum islands. The white arrows indicate a monolayer graphene step and the dashed line locates the position of the spectroscopy measurements in Figs. 2 and 3. (c) High resolution STM image of the graphene lattice obtained at the location marked by the red square in (b). The larger modulations are due to a moiré pattern with period of ≈2.8 nm due to rotational misalignment of ≈5° between the top and second graphene layers.

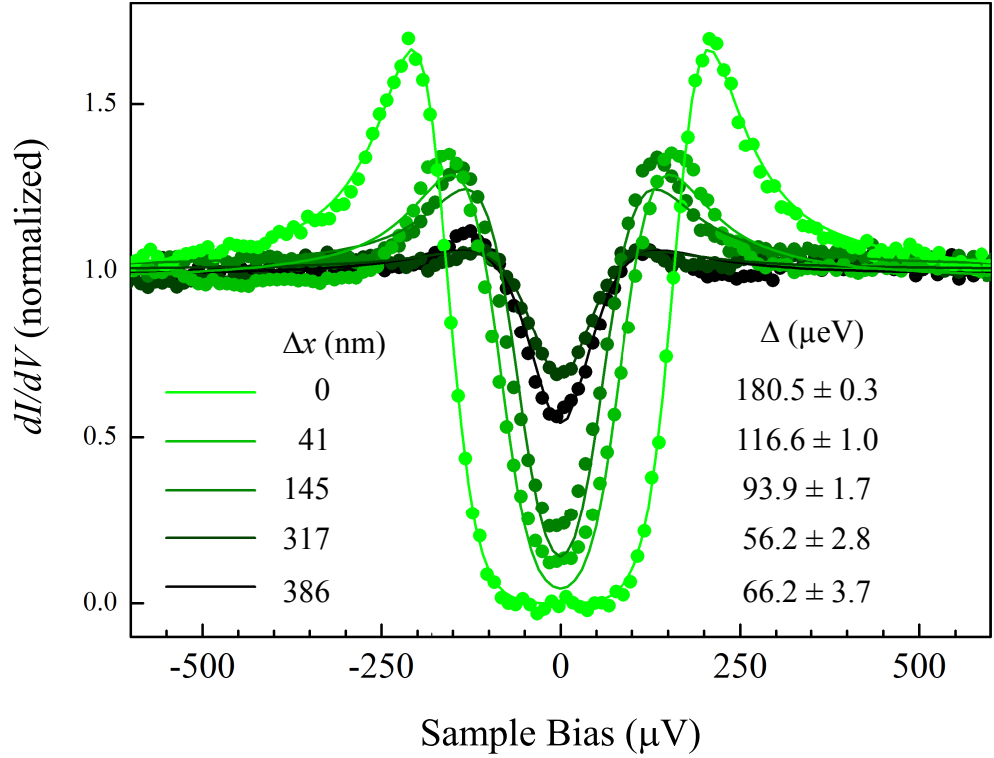

Fig. 2. Graphene superconducting tunneling spectroscopy. Differential tunneling spectra (symbols) measured at several lateral positions from an aluminum-graphene edge at $\Delta x$=0 (see Fig. 3). The solid lines are non-linear fits using the modified BCS theory by Maki [19,20] with superconducting gaps indicated on the right of the graph [21]. An effective temperature of $T_{eff}$=232 mK representing the residual electrical noise in the system, was determined from the best fit to the aluminum spectrum at $x$=0 and subsequently held fixed to extract the distance dependent gap width on graphene. The error in the gap energy was determined from the chi-square minimization in non-linear least square fits to the Maki theory.

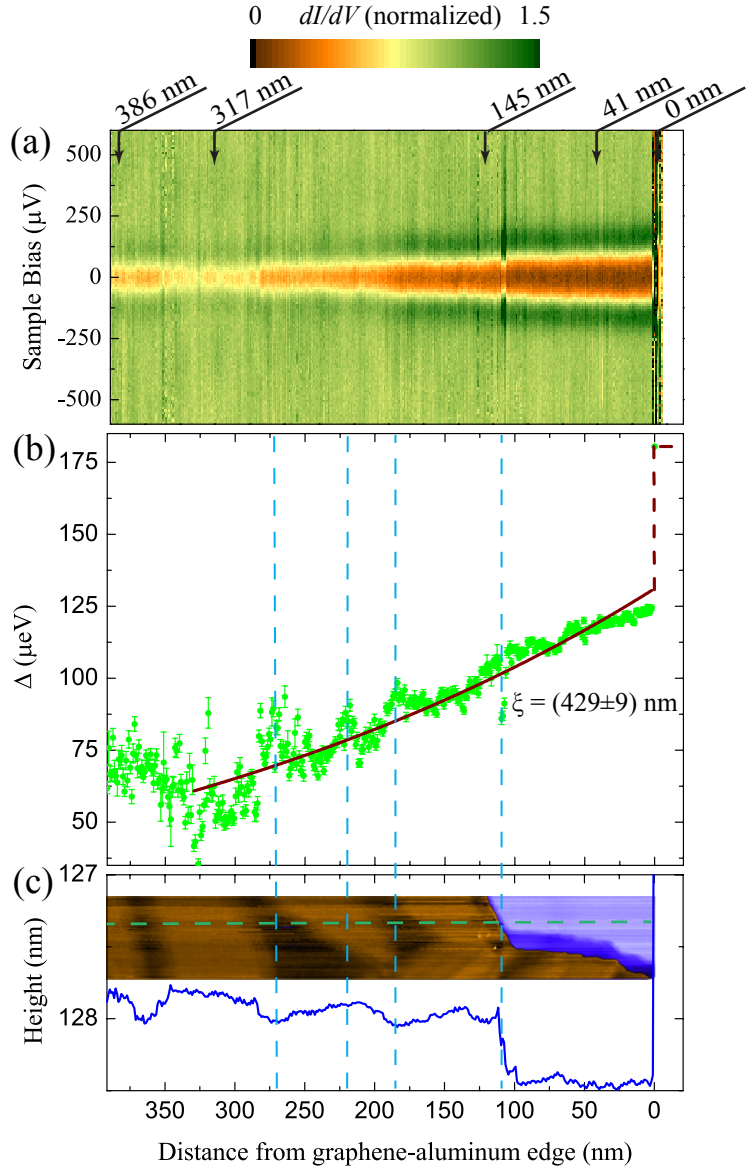

Fig. 3. Proximity induced superconductivity in epitaxial graphene. (a) *dI/dV* vs $V_b$ tunneling spectra, measured on the graphene terrace starting at a graphene-aluminum edge along the dashed lines shown in (c) and Fig. 1(b). The spectra are displayed in a color scale, where the brown color indicates a superconducting gap induced by proximity to the nearby aluminum islands. (b) The superconducting gap [21] determined by fitting the spectra in (a) to the modified BCS theory of Maki [19,20]. The dashed line is a guide to the eye to show the abrupt change in gap energy near the graphene-aluminum interface. The gap energies for (325 nm≥x>0 nm) are fit to an exponential decay yielding a graphene coherence length of $\xi = (429\pm9)$ nm [21]. (c) The graphene topographic height and STM image along the path of the spectral measurements in (a) and (b).